\documentclass[aps,prb,superscriptaddress,amsfonts,amsmath,amssymb,showpacs,floatfix,reprint,longbibliography]{revtex4-1}

\usepackage{url}
\usepackage{bm}
\usepackage{graphicx}
\usepackage{amsmath}
\usepackage{amstext}
\usepackage{amssymb}
\usepackage{amsfonts}
\usepackage{amsbsy}
\usepackage{verbatim}
\usepackage{color}
\usepackage[colorlinks=true, urlcolor=blue, linkcolor=blue, citecolor=blue, pdftex]{hyperref}
\usepackage{multirow}
\usepackage{floatrow}
\usepackage{float}
\usepackage{gensymb}
\usepackage{textcomp}
\usepackage{enumitem}
\usepackage[version=3]{mhchem}
\usepackage{afterpage}
\usepackage{pbox}
\usepackage{makecell}
\usepackage{array,booktabs}
\usepackage{dsfont}
\usepackage{siunitx}

\usepackage{bbold} 
\usepackage{color}
\definecolor{notecolor}{rgb}{0.0,0.2,0.8}
\definecolor{karlocolor}{rgb}{0.9,0.2,0}
\definecolor{pocokcolor}{rgb}{0.9,0.05,0.05}

\begin{document}

\title{Affine lattice construction of spiral surfaces in classical Heisenberg models}

\author{P\'eter Balla}
\affiliation{Institute for Solid State Physics and Optics, Wigner Research Centre for Physics, Hungarian Academy of Sciences, H-1525 Budapest, P.O.B. 49, Hungary}
\author{Yasir Iqbal}
\affiliation{Department of Physics, Indian Institute of Technology Madras, Chennai 600036, India}
\author{Karlo Penc}
\affiliation{Institute for Solid State Physics and Optics, Wigner Research Centre for Physics, Hungarian Academy of Sciences, H-1525 Budapest, P.O.B. 49, Hungary}

\date{\today}

\begin{abstract}
Frustration in classical spin models can lead to degenerate ground states without long range order. In reciprocal space, these degeneracies appear as manifolds of wave vectors, their dimensionality increasing with the degree of frustration and the robustness of the disordered spin-liquid state. Here, we present a recipe to explicitly construct Heisenberg models on Bravais lattices with codimension-one manifolds, i.e., lines in two-dimensions and surfaces with different Euler characteristics in three-dimensions. Furthermore, we discuss the role of thermal and quantum fluctuations in stabilizing ordered states.
\end{abstract}

%\pacs{75.10.-b, 75.10.Jm, 75.10.Pq, 75.30.Kz, 75.40.Mg}

\maketitle

%%%%%%%%%%%%%%%%%%%%%%%%%%%

 Many an endeavor in the modern era of quantum magnetism has centered around finding exciting escape routes from the seemingly inevitable fate that befalls an overwhelming majority of magnetic systems; namely, spontaneous symmetry breaking at low temperatures and the consequent development of long-range magnetic order. The lure is to find exotic phases of matter called spin liquids\textemdash states which lack a local order parameter down to zero-temperature and thus lie beyond the realm of Landau's symmetry breaking theory~\cite{Landau-1937a,*Landau-1937b}. Spin liquids occur in two genres, (i) quantum spin liquids~\cite{anderson73,balents10}\textemdash featuring complex patterns of long-range entanglement, quasiparticles with fractional quantum numbers and possibly nonabelian statistics and (ii) classical spin liquids~\cite{ramirez99,bramwell01}\textemdash cooperative paramagnetic states of classical ($S\to\infty$) spins featuring nontrivial spin correlations~\cite{Moessner-1998a,*Moessner-1998b}, and for certain types, fractionalization~\cite{Rehn-2017}. 
 
 The traditional route towards finding quantum spin liquids involves melting magnetic order via strong quantum fluctuations and preferably occurs in models combing low spin with geometrically and/or parametrically frustrated interactions. 
 In the complete absence of quantum fluctuations, as for classical spins, the quenching of magnetic order is, nevertheless possible, but now crucially hinges on the existence of a macroscopic degeneracy of the ground state manifold $\mathcal{M}_{\textrm{GS}}$ within which the system fluctuates in a cooperative fashion giving rise to the notion of a classical spin liquid~\cite{Moessner-1998a,Moessner-1998b,bergmann07,gao16}. A macroscopic degeneracy can emerge in two possible scenarios: (i) The presence of local ice-rule type constraints~\cite{pauling35} which define the set of allowed ground states but leave the ground-state spin configurations underdetermined. This situation occurs, e.g., in the Heisenberg antiferromagnet on the pyrochlore lattice, wherein, the zero magnetization per tetrahedron constraint gives rise to an extensively degenerate $\mathcal{M}_{\textrm{GS}}$~\cite{Moessner-1998a,Moessner-1998b} and (ii) If  $\mathcal{M}_{\textrm{GS}}$ is composed of a highly degenerate family of incommensurate coplanar spin-spirals. This situation is realized in the Heisenberg antiferromagnetic model with first and second nearest-neighbor interactions on the honeycomb~\cite{Katsura1986,fouet01,mulder10,baez17,Shimokawa2018} and diamond lattices~\cite{bergmann07,Attig2017,Iqbal-2018}. The spiral wave vectors {\bf Q} form, in the former case, contours, and in the latter, a closed surface in reciprocal space. 
 
 The existence of a macroscopic degeneracy although being a necessary ingredient to realize classical spin liquids is by no means sufficient. Indeed, only under the condition that thermal \emph{order-by-disorder} effects fail to lift this degeneracy and select a unique ground state, does one realize a true classical spin liquid as a zero-temperature phase. However, in the scenario (ii) even if thermal order-by-disorder mechanism leads to magnetic ordering (at a particular wave vector $\mathbf{Q}$) at a temperature $T_{c}$, there exists a temperature window above $T_{c}$ and below the Curie-Weiss temperature in which thermal fluctuations can restore the spiral surface~\cite{bergmann07,gao16,Iqbal-2018}. Within this cooperative paramagnetic regime the spins engage in collective motion within this spiral manifold leading to the appearance of a finite-temperature spiral spin liquid. Given this wealth of phenomena that can potentially emerge from the presence of a spiral surface, our work provides recipes for constructing frustrated classical Heisenberg models on the simplest lattices, namely the \emph{Bravais lattices} which host a spin spiral surface. We will illustrate our method on a few frequently encountered lattices, namely on the square, simple cubic (SC), and face-centered cubic (FCC) ones.   

The classical isotropic Heisenberg model is defined by the Hamiltonian
\begin{eqnarray}
&&\mathcal{H}=\frac{1}{2}\sum_{i,\bm{\delta}}J_{\delta}\mathbf{S}_{i} \cdot \mathbf{S}_{i+\bm{\delta}} \;,
\label{eq:Hamilton}
\end{eqnarray}
where $\mathbf{S}_{i}$ are three-dimensional unit vectors at the sites $\mathbf{R}_{i}$ of a Bravais lattice $\Lambda$, with $N$ sites and periodic boundary conditions. The  $J_{\delta}$ are the exchange couplings between spins at sites separated  by $\bm{\delta}$, the neighbor vectors of $\Lambda$. The $J_{\delta}$ can be sorted by the increasing norm of $\bm{\delta}$, and hence it is convenient to adopt the notation wherein $J_1$, $J_{2}$, $J_{3}$, $\ldots$
%, $J_{n}$ 
denote first-, second-, third-, $\ldots$ 
%$n$-th 
nearest-neighbor exchange couplings, respectively.

In the spirit of the Luttinger-Tisza method \cite{luttinger_tisza,*luttinger_tisza1}, the ground state of the model [Eq.~\eqref{eq:Hamilton}] can be found by minimizing the Fourier transform $J(\mathbf{q})$ of the exchange interactions. The energy in reciprocal space is given by
\begin{equation}
\mathcal{H} = \frac{N}{2}\sum_{\mathbf{q}\in\textrm{BZ}}
  J(\mathbf{q})\mathbf{S}_{\mathbf{q}} \cdot \mathbf{S}_{-\mathbf{q}}\;,
\label{eq:Hamilton_FT}\\
\end{equation}
where the summation runs over the Brillouin-zone (BZ), 
\begin{equation}
J(\mathbf{q}) = \sum_{\bm{\delta}} J_{\delta} e^{\imath\mathbf{q}\cdot\bm{\delta}}  \;,
\label{eq:jq_def}
\end{equation}
and $\mathbf{S}_{\mathbf{q}}=\frac{1}{N}\sum_{i} \mathbf{S}_{i} e^{\imath\mathbf{q}\cdot\mathbf{R}_i}$. We denote the ground state manifold by $\mathcal{M}_{\textrm{GS}}=\left\{\mathbf{Q}\right\}$, the set of points where $J(\mathbf{q})$ takes its minimal value. Generally $\mathcal{M}_{\textrm{GS}}$ is only a set of discrete points, and they  correspond to the ordering vectors of the ground-state spin configuration $\mathbf{S}_i=\sum_{\mathbf{Q}}\mathbf{S}_{\mathbf{Q}} e^{-\imath\mathbf{Q}\cdot\mathbf{R}_i}$. The  constraint $\left|\mathbf{S}_{i}\right|=1$ can usually be satisfied by a proper selection of complex amplitudes $\mathbf{S}_{\mathbf{Q}}$~\cite{villain_rules,henley_fcc,yamamoto1972,nagamiya_helical_book_chapter,Nussinov2001}. In particular, models are known in which $\mathcal{M}_{\textrm{GS}}$ consist of lines~\cite{alexander1980,mulder10}, surfaces~\cite{finnish_93,bergmann07,Attig2017}, or even the complete BZ for certain non-Bravais lattices~\cite{Reimers-1992,Iqbal-2019}.  
The dimensionality of $\mathcal{M}_{\textrm{GS}}$ is known to be a crucial ingredient in determining the physical behavior of the model, with higher dimensionality being associated with increased frustration. Here, we present  a systematic method to construct exchange models $J(\mathbf{q})$ that have codimension-one $\mathcal{M}_{\textrm{GS}}$'s, {\it i.e.}, a curve for a 2D lattice or a surface for a 3D lattice. 
%%%%%%%%%%%%%%%%%%%%%%%%%%%%%%%%%%%%%%
% Affine lattice construction
%%%%%%%%%%%%%%%%%%%%%%%%%%%%%%%%%%%%%%

To ensure that the minimum of $J(\mathbf{q})$ is on a  codimension-one $\mathcal{M}_{\textrm{GS}}$  defined by $f(\mathbf{Q}) = 0$, we make the following Ansatz:
\begin{equation}
  J(\mathbf{q}) = f^2(\mathbf{q}) - C, \quad f(\mathbf{q}) = \sum_{\bm{\xi}\in \Xi} c_{\bm{\xi}} e^{\imath \mathbf{q} \cdot\bm{\xi}} \;,
   \label{eq:Jq=fq2}
\end{equation}
where $\Xi$ is a set of points in real space, and $f(\mathbf{q})$ is a real function of the wave-vector 
\footnote{A complex $f(\mathbf{q}) = f'(\mathbf{q}) + \imath f''(\mathbf{q})$ in a $J(\mathbf{q})= \left|f(\mathbf{q})\right|^2-C$ would result in a codimension-two $\mathcal{M}_{\textrm{GS}}$ defined by $f'(\mathbf{q}) = 0$ and $ f''(\mathbf{q}) = 0$}. 
$J(\mathbf{q})$ then takes the form 
\begin{equation}
  J(\mathbf{q}) = \sum_{\bm{\xi},\bm{\xi}' \in \Xi} c_{\bm{\xi}} c_{\bm{\xi}'} e^{\imath \mathbf{q} \cdot (\bm{\xi} + \bm{\xi}' )} - C. 
  \label{eq:jqf2p}
 \end{equation}
What are the possible choices of the $\Xi$ sets, for which the  $J(\mathbf{q})$ above provides a genuine Heisenberg model in Fourier space, as defined in  Eq.~(\ref{eq:jq_def})?
 Since the reality of $f(\mathbf{q})$ requires that for any $\bm{\xi} \in  \Xi$, the $-\bm{\xi}$ is also in $\Xi$, and $c_{-\bm{\xi}} = c_{\bm{\xi}}^*$, we may substitute $\bm{\xi}'\rightarrow -\bm{\xi}'$ in Eq.~(\ref{eq:jqf2p}) yielding $\bm{\xi} - \bm{\xi}' \in \Lambda$. Therefore, $\Xi$ is, by definition, an inversion symmetric finite subset of an {\it affine lattice} $\Lambda^*$, {\it i.e.},
 $\Lambda$ shifted by some vector $\bm{\delta}^*$: 
	\begin{equation}
	\Lambda^* = \Lambda + \bm{\delta}^*, \textrm{ with }2 \bm{\delta}^* \in \Lambda \,, 
	\label{eq:2delta}
	\end{equation}
where the condition for  $\bm{\delta}^*$ follows from Eq.~(\ref{eq:jqf2p}) by substituting $\bm{\xi}'\rightarrow \bm{\xi}$. There are four choices for $\Lambda^*$ in two dimensions and eight in three dimensions. 

%So far we have only used  the translational symmetries of $\Lambda$. 
 To get a model having the full symmetry of $\Lambda$, the site symmetry group of  $\bm{\delta}^*$ (or any other point in $\Lambda^*$) needs to be isomorphic to the point group $\mathcal{G}$ of $\Lambda$. While 
 $\bm{\delta}^* =\mathbf{0}$ gives a correct model for every Bravais lattice, the site symmetry of a $\bm{\delta}^*\neq \mathbf{0}$ can be looked up in crystallographic tables \cite{Bilbao}, and it turns out that  such $\bm{\delta}^*$  exists for all Bravais lattices except for the triangular lattice in 2D and body-centered cubic lattice in 3D.
 
 To also utilize the point symmetry  $\mathcal{G}$ in our Ansatz, we first decompose $\Lambda^*$ into orbits (shells) $\Xi_\alpha$  under the action of $\mathcal{G}$: for a $\bm{\delta}^*_\alpha \in \Lambda^*$ let $\Xi_\alpha \colon =\left\{g \bm{\delta}^*_\alpha | \ g\in \mathcal{G} \right\}$, the index $\alpha$ enumerates the shells.  
 Based on an orbit we define  symmetry adapted functions \cite{tinkham}:
\begin{equation}
  f_\alpha^\Gamma(\mathbf{q})= \sum_{g\in \mathcal{G}}\chi^\Gamma ( g )  e^{\imath \mathbf{q} \cdot(g\bm{\delta}_\alpha^*)}\;,
  \label{eq:fsymm0}
\end{equation}
where $\chi^\Gamma ( g )=\pm 1$ are the characters of a 1D real irreducible representation $\Gamma$. 
We  then choose a set of orbits $\Xi_\alpha$ and corresponding constants $c_\alpha\in \mathds{R}$ or $i\mathds{R}$ to get the real
    \begin{equation}
   f^\Gamma(\mathbf{q})= \sum_{\alpha \in \textrm{orbits}} c_\alpha f_\alpha^\Gamma(\mathbf{q})\;.
   \label{eq:fsymm}
   \end{equation}
From Eq.~(\ref{eq:Jq=fq2}) with $C= \sum_{\alpha} z_\alpha |c_\alpha|^2 $, where $z_\alpha= |\Xi_\alpha|$ and $\Xi=\bigcup_\alpha \Xi_\alpha$,  we get a Heisenberg model with a   $\mathcal{M}_{\textrm{GS}}$ defined by $f^\Gamma(\mathbf{q})= 0$. The number of free parameters as well as the range of exchange couplings grows with the number of shells in  $f^\Gamma(\mathbf{q})$. In what follows we will only use the totally symmetric representation ($\chi^\Gamma(g)=1$) of $\mathcal{G}$, and  therefore drop the index $\Gamma$.

\begin{figure}[t]
	\centering
	    \includegraphics[width=0.85\columnwidth]{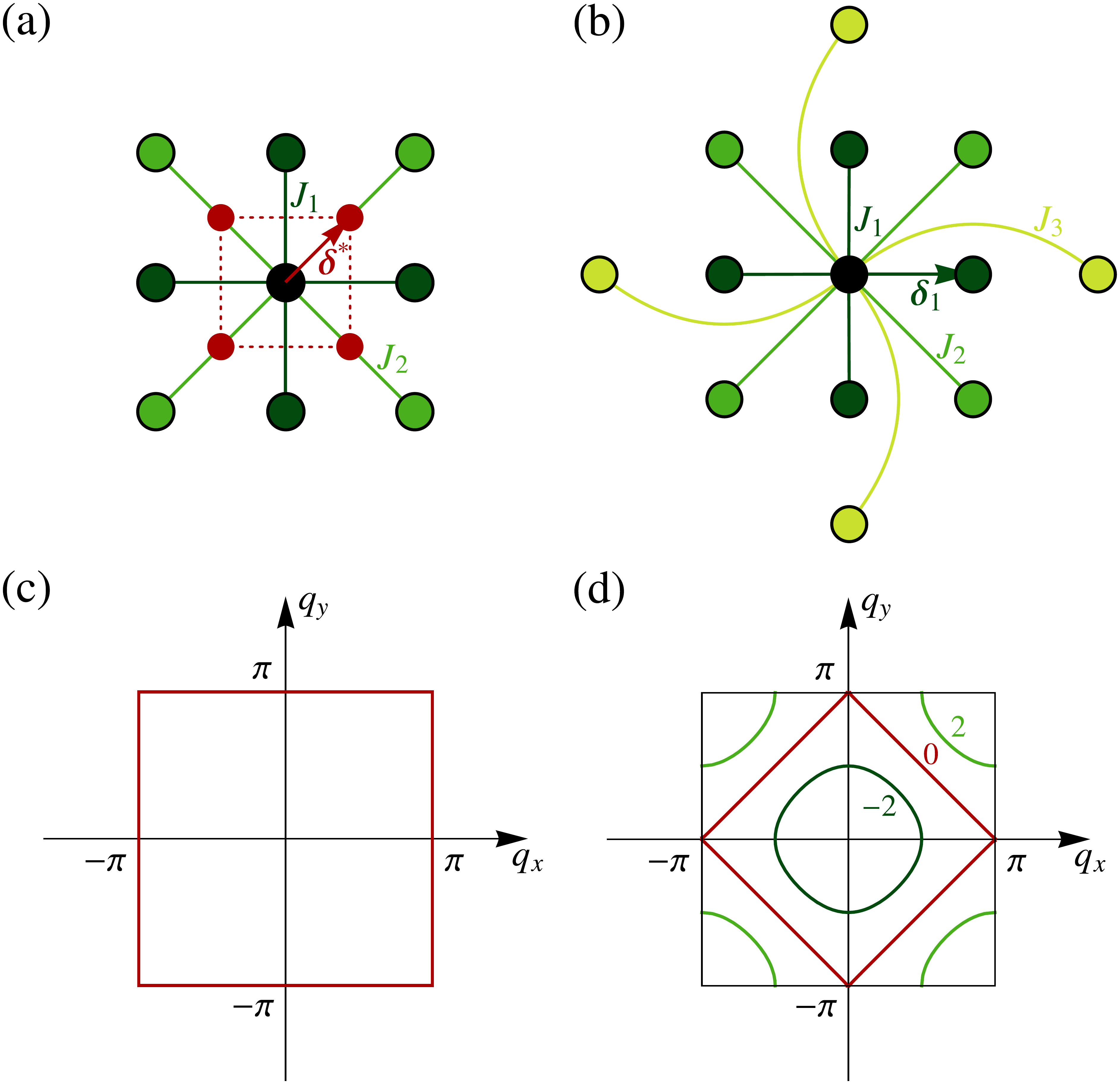}
	\caption{
	 Constructions of the codimension-one ground state manifold $\mathcal{M}_{\textrm{GS}}$, given by the minima of    $J(\mathbf{q})$ defined by $f(\mathbf{q})=0$ via Eq.~(\ref{eq:Jq=fq2}), on the square lattice $\Lambda$.
	(a) Affine lattice construction with $\Lambda^* = \Lambda+\bm{\delta}_1^*$. The red arrows connect the origin with the points of set $\Xi_1 \subset \Lambda^*$ (red dots), dark green dots are the nearest-neighbor points to the origin (black dot) of $\Lambda$ with exchange coupling $J_1$, light green dots show the second nearest-neighbor points with $J_2=J_1/2$.
	(b) Construction when  $\Xi_0, \Xi_1 \subset \Lambda$, the $\Xi_0$ has one point, the origin (black dot) and the $\Xi_1$ is the orbit of $\bm{\delta}_1=\mathbf{a}_1$ (dark green dots). We get a Heisenberg model with nearest-neighbor exchanges $J_1$ (dark green) and further neighbor exchanges  $J_3=J_2/2$ (denoted by lighter colors). 
	 (c) $\mathcal{M}_{\textrm{GS}}$ (red square) for $\Lambda^* = \Lambda+\bm{\delta}_1^*$, which corresponds to the zeros of the function $f(\mathbf{q})$ in Eq.~\eqref{eq:squ_dual_f}, is pinned to the Brillouin zone boundary. 
	(d) $\mathcal{M}_{\textrm{GS}}$ for $\Lambda^* = \Lambda$, given by $f(\mathbf{q})=0$ in Eq.~\eqref{eq:epsfromgq}, are shown as thick colored curves for $J_1/J_2=2$, $0$, and $-2$.
	}
	\label{fig:square_lattice}
\end{figure}

As an illustration, we show how the method works for the square lattice $\Lambda$ with primitive vectors $\mathbf{a}_1=\left( 1,0\right)$ and $\mathbf{a}_2=\left(0, 1\right)$, shown in Fig.~\ref{fig:square_lattice}. Out of the four choices for $\bm{\delta}^*$,
only the cases when $\bm{\delta}^* = \mathbf{0}$ and $\bm{\delta}^* =\left(\frac{1}{2},\frac{1}{2}\right)$ have $\mathcal{G}=\mathsf{D_4}$ as the site symmetry group.
 
First, let us consider the $\bm{\delta}^* =\left(\frac{1}{2},\frac{1}{2}\right)$ case [Fig.~\ref{fig:square_lattice}(a)]. We choose the first shell $\Xi=\Xi_1$ with cardinality $z_1=4$ and consisting of the orbit of 
 $\bm{\delta}_1^*=\bm{\delta}^*$.   Eqs.~(\ref{eq:fsymm0}) and (\ref{eq:fsymm}) give
\begin{equation}
f(\mathbf{q}) = f_1(\mathbf{q})=4\cos \frac{q_x}{2}\cos \frac{q_y}{2},
\label{eq:squ_dual_f}
\end{equation}
and the $\mathcal{M}_{\textrm{GS}}$ coincides with the BZ boundary $\mathbf{Q} = (\pi,q)$ and $(q,\pi)$ parametrized by $q\in [-\pi,\pi]$,  [see Fig.~\ref{fig:square_lattice}(c)]. 
Following Eq.~\eqref{eq:Jq=fq2},
\begin{align}
J(\mathbf{q}) &= 4(\cos q_x+\cos q_y) \nonumber \\
  &\phantom{=}+ 2\left[\cos \left(q_x + q_y\right)+\cos \left(q_x - q_y\right)\right]
\end{align}
defines a $J_{1}$\textendash$J_{2}$ Heisenberg model with exchange couplings $J_{1}=2$ and $J_{2}=1$ \cite{ChandraJ1-J2_1988}.   
The Hamiltonian is the sum of edge sharing four-site complete graphs (squares with diagonals) over the lattice $\Lambda$: 
\begin{equation}
\mathcal{H}=\sum_{\boxtimes}{\left[\left(\mathbf{S}_1+\mathbf{S}_2+\mathbf{S}_3+\mathbf{S}_4\right)^2- 4 \right]},
\label{eq:tetra_cover}
\end{equation}
which is minimized when the spins sum up to zero in every graph. 
We note that an alternative approach to construct this $\mathcal{M}_{\textrm{GS}}$ was presented in Ref.~\cite{Attig2017}.

When $\bm{\delta}^* = \mathbf{0}$,  $\Lambda^*=\Lambda$, and we choose $\Xi=\Xi_0\cup\Xi_1$, where $\Xi_0=\{\mathbf{0}\}$ and $\Xi_1=\{\mathbf{a}_1,\mathbf{a}_2,-\mathbf{a}_1,-\mathbf{a}_2\}$, with $z_0=1$ and $z_1=4$, to construct  
\begin{equation}
f(\mathbf{q})=  f_1(\mathbf{q}) + c_0 = 2(\cos q_x+\cos q_y)+c_0
\label{eq:epsfromgq}
\end{equation} 
following Eq.~(\ref{eq:fsymm}) with  $c_1=1$. 
Eq.~\eqref{eq:Jq=fq2} then generates a 
%$J_1$\textendash$J_2$\textendash$J_3$ 
model with $J_1= 2c_0$, $J_2=2$, and a constrained $J_3=J_2/2$ [see Fig.~\ref{fig:square_lattice}(b)], also discussed in \cite{niggemann_arxiv}. By tuning the parameter $c_0=J_1/J_2$ one can control the shape and topology of  $\mathcal{M}_{\textrm{GS}}$, as shown in Fig.~\ref{fig:square_lattice}(d).

%%%%%%%%%%%%%%%%%%%%%%%%%%%%%%%%%%%%%%%%%%%%%%%%%%%%%%%%%%%%%%%%%%
%\subsection{3D examples: SC and FCC, free energies}
%%%%%%%%%%%%%%%%%%%%%%%%%%%%%%%%%%%%%%%%%%%%%%%%%%%%%%%%%%%%%%%%%%

In what follows, we construct models for the SC and FCC lattices based on the affine lattice construction $\Lambda^*$ with $\bm{\delta}^*\neq \mathbf 0$, both having $\mathcal{G}=\mathsf{O_h}$ as a point group, and calculate their free energies on the resulting $\mathcal{M}_{\textrm{GS}}$-s in order to find the states stabilized by thermal and quantum fluctuations \cite{bergmann07}. 

\begin{figure}[h]
\centering
\includegraphics[width=0.95\columnwidth]{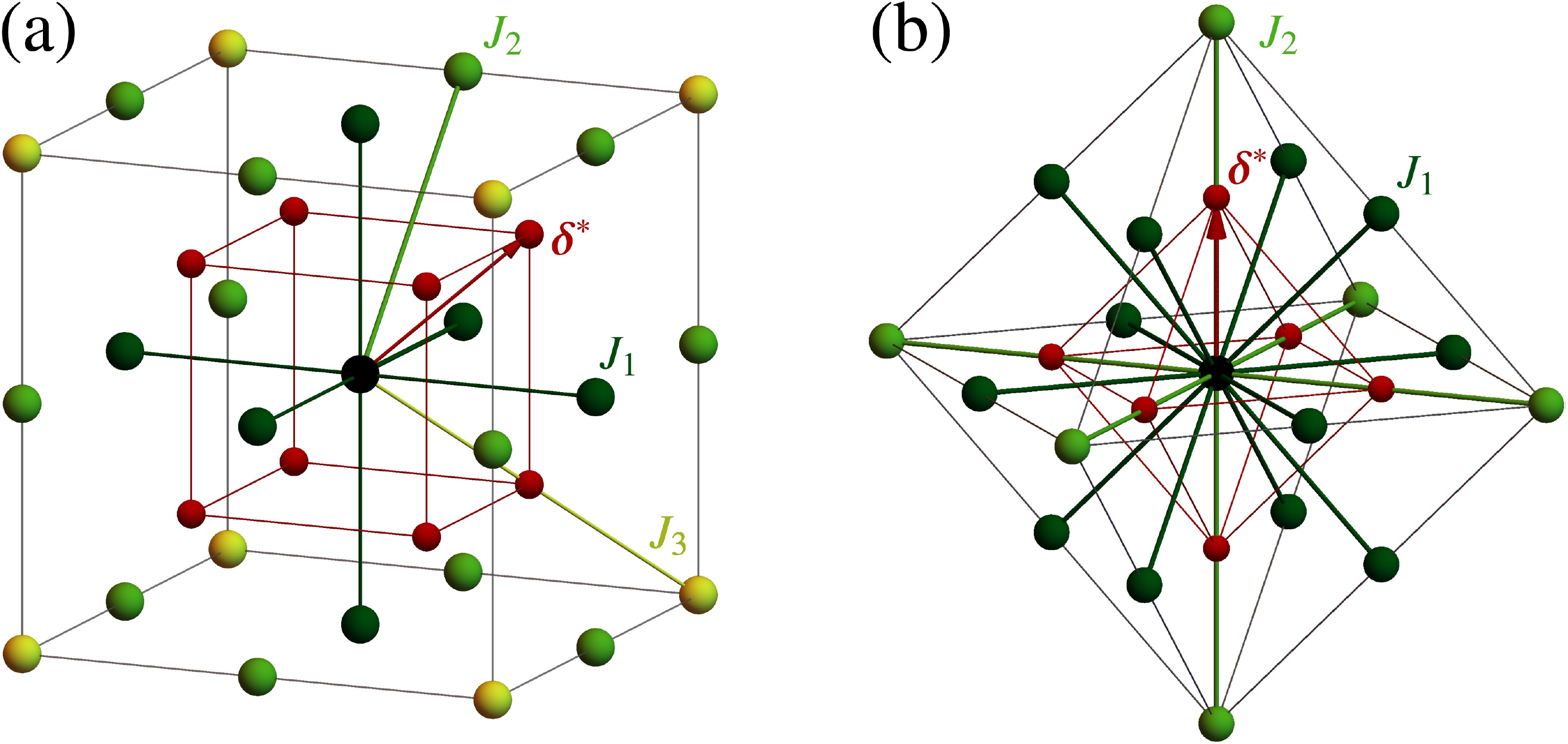}
\caption{Affine lattice constructions for the simple cubic (a) and face centered cubic (b) lattices. The shift vector $\bm{\delta}^*$ (red arrow) defines the affine lattice $ \Lambda^*=\Lambda+\bm{\delta}^*$. Red balls with a cage show the set $\Xi_1\subset \Lambda^*$. Dark balls are the nearest neighbor points to the origin (black ball) of  $\Lambda$, with exchange couplings $J_1$. Lighter balls depict the second neighbor points with exchange strengths $J_2=J_1/2$ for both lattices. For the simple cubic lattice a third neighbor exchange $J_3=J_2/2$ is also generated.}
\label{fig:SC_FCC_dualconstruct}
\end{figure} 

\begin{figure}[h]
	\centering
		\includegraphics[width=0.8\columnwidth]{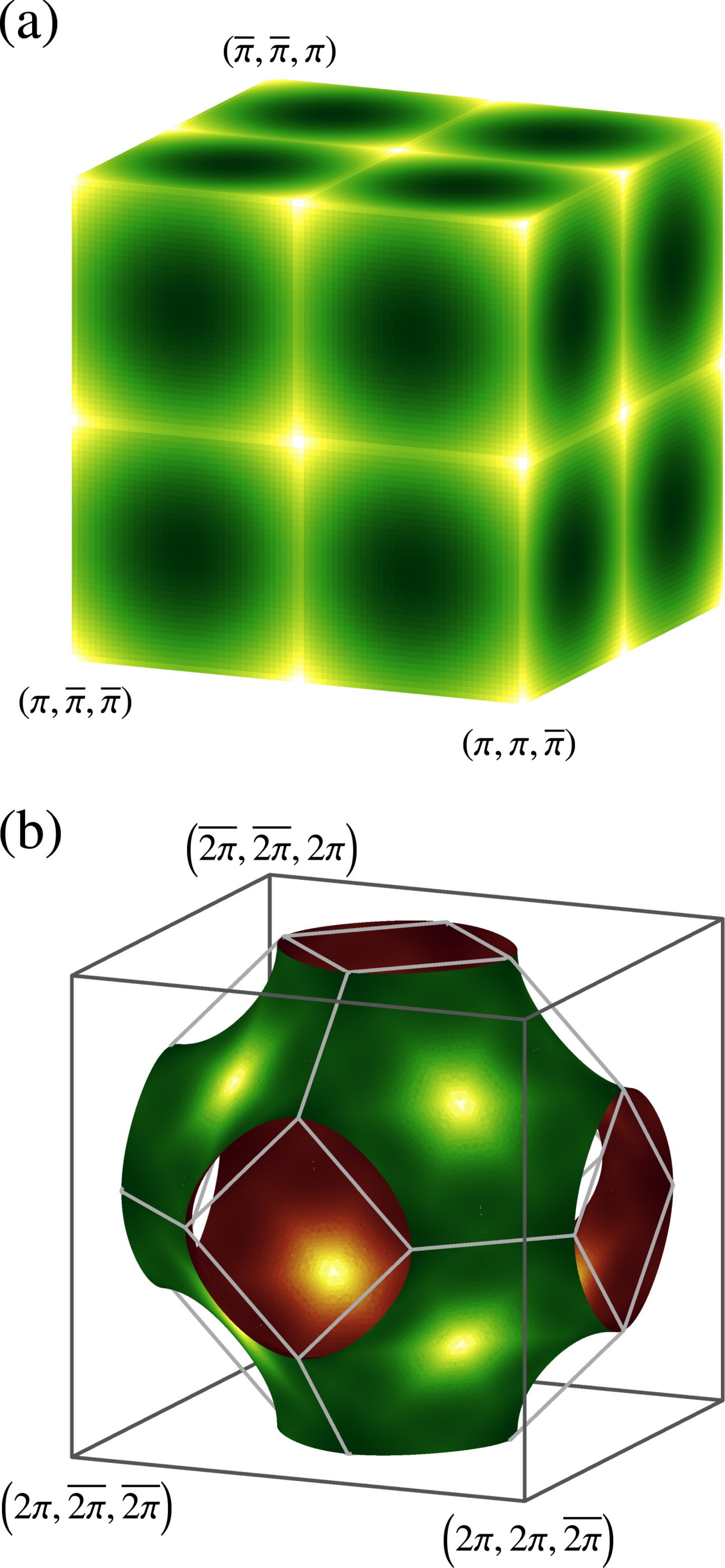}
	\caption{Two examples of the ground state manifolds of 3D lattices colored according to their free energies for the affine lattice construction with $\bm{\delta}^*\neq\mathbf{0}$. Brighter colors correspond to states with smaller values of $\mathcal{A}\left(\mathbf{Q}\right)$ [Eq.~\eqref{eq:entropy}], and the minima are selected by thermal fluctuations. (a) Simple cubic lattice: the ground state manifold is the Brillouin zone boundary, the degenerate minima are the inequivalent points $\left(\pi,0,0\right)$, $\left(\pi,\pi,0\right)$ and $\left(\pi,\pi,\pi\right)$ and their symmetry related partners. (b) Face centered cubic lattice: the degenerate minima are the points $\langle \pi,\pi,\pi\rangle$. The Brilloin zone boundary is shown as a light wireframe, the enclosing cube is a guide to the eye.
	 }
	\label{fig:dual_free_energies}
\end{figure}

The SC lattice is defined by the primitive vectors  $\left(1,0,0\right)$, $\left(0,1,0\right)$ and $\left(0,0,1\right)$. In the affine construction,
we start with the orbit of the vector $\bm{\delta}^*=\left(\frac{1}{2},\frac{1}{2},\frac{1}{2}\right)$,  and from Eqs.~(\ref{eq:fsymm0}) and (\ref{eq:fsymm}) we get
\begin{equation}
f(\mathbf{q})=8\cos \frac{q_x}{2}\cos \frac{q_y}{2}\cos \frac{q_z}{2}.
\label{eq:f_SC}
\end{equation}
This generates a $J_1$\textendash$J_2$\textendash$J_3$ model [see Fig.~\ref{fig:SC_FCC_dualconstruct}(a)], with  $J_3=J_2/2=J_1/4=1$. The resulting $\mathcal{M}_{\textrm{GS}}$ is the cubic BZ boundary shown in Fig.~\ref{fig:dual_free_energies}(a). As in Eq.~\eqref{eq:tetra_cover}, the Hamiltonian is the sum of face sharing eight-site complete graphs (elementary cubes with face and body diagonals),
which is minimized when the spins sum up to zero in every graph. In comparison, the construction with $\bm{\delta}^* = \mathbf{0}$ and two shells gives  $J_4=J_2/2=1$ and an adjustable $J_1=2  c_0$.

The face centering generators of the FCC lattice are 
$\left(\frac{1}{2},\frac{1}{2},0\right)$, $\left(\frac{1}{2},0,\frac{1}{2}\right)$ and $\left(0,\frac{1}{2},\frac{1}{2}\right)$. In the affine construction $\bm{\delta}^*=\left(0,0,\frac{1}{2}\right)$ and $\Xi_1 = \left\{ \left(\pm\frac{1}{2},0,0\right), \left(0,\pm\frac{1}{2},0\right), \left(0,0,\pm\frac{1}{2}\right)\right\}$ with $z_1=6$, such that  
\begin{equation}
f_1(\mathbf{q})=2\left(\cos \frac{q_x}{2}+\cos \frac{q_y}{2}+\cos\frac{q_z}{2}\right).
  \label{eq:fgfcc} 
\end{equation}
We get a $J_1$\textendash$J_2$ model with $J_2=J_1/2=1$ [see Fig.~\ref{fig:SC_FCC_dualconstruct}(b)]. 
This construction provides  $\mathcal{M}_{\textrm{GS}}$ studied in Ref.~\cite{finnish_93} and shown in Fig.~\ref{fig:dual_free_energies}(b). Similar to Eq.~\eqref{eq:tetra_cover}, the Hamiltonian is the sum of edge sharing six-site complete graphs (octahedra with diagonals).
In contrast, the construction with $\bm{\delta}^*=\mathbf{0}$ provides a model with $J_4=J_3/2=J_2/4=1$ and adjustable $J_1=4  + 2 c_0$, and the minimal energy surface is the same as for the diamond lattice \cite{bergmann07}. 

%%%%%%%%%%%%%%%%%%%%%%%%%%%%%%%%%%%%
%Free energy part
%%%%%%%%%%%%%%%%%%%%%%%%%%%%%%%%%%%%
%

At zero temperature any of the different $\mathbf{Q}$ can be selected as a good ground state, and even multiple $\mathbf{Q}$ states are allowed at points with high symmetry \cite{villain_rules}. At finite temperature the spins start to fluctuate, and this contributes to the explicitly $\mathbf{Q}$-dependent free energy $\mathcal{F}(\mathbf{Q})$. In the harmonic approximation, the fluctuations can be integrated out in the partition sum, and give rise to a linear $T$ dependence in $\mathcal{F}(\mathbf{Q})$. Following Ref.~[\onlinecite{bergmann07}], the free energy above the spiral surface is
\begin{equation}
\mathcal{F}(\mathbf{Q}) = \mathcal{E}_0 - N T \ln T +  N T \mathcal{A}(\mathbf{Q}) + \mathcal{O}(T^2) \;,
\label{eq:free_energ}
\end{equation}
where $\mathcal{E}_0$ is the ground state energy and $-\mathcal{A}(\mathbf{Q})$ is the $\mathbf{Q}$-dependent part of the low temperature entropy density, defined as
\begin{equation}
 \mathcal{A}(\mathbf{Q}) = \frac{1}{N}\sum_{\mathbf{q} \in \textrm{BZ}} \ln \omega_{\mathbf{q}}^2(\mathbf{Q})  ,
\label{eq:entropy}
\end{equation}
with
\begin{align}
 \omega_\mathbf{q}(\mathbf{Q}) &= \left( \frac{1}{2}\left[ J(\mathbf{q}\!+\!\mathbf{Q})+J(\mathbf{q}\!-\!\mathbf{Q}) - 2 J(\mathbf{Q})\right]  \right. \nonumber\\
  &\left. \vphantom{\frac{1}{2}}\phantom{=}\times \left[ J(\mathbf{q}) - J(\mathbf{Q}) \right] \right)^{1/2} \;.
\label{eq:omega}
\end{align}
The state which has the minimal value of $\mathcal{A}(\mathbf{Q}) $ corresponds to the minimum of the free energy --- this is the entropic order-by-disorder selection mechanism discussed in Refs.~[\onlinecite{villain1980,shender1982,Kawamura1984,henley1989}]. 
We plot $\mathcal{A}(\mathbf{Q})$ for the SC and FCC lattice in Fig.~\ref{fig:dual_free_energies}. 
Furthermore, the energy of the spin-wave modes is $ \hbar S\omega_\mathbf{q}(\mathbf{Q})$  in the semi-classical description for spins of length $S\gg 1$. {\it Quantum} fluctuations then choose the state with the lowest zero-point energy 
%We note, that for quantum spins of length $S$ in the semi-classical description, i.e., $1/S\ll 1$, the energy of the spin-wave modes is $ \hbar S\omega_\mathbf{q}(\mathbf{Q})$, and {\it quantum} fluctuations choose the state with the lowest zero-point energy 
%
\begin{equation}
 \mathcal{E}_{\text{ZP}}(\mathbf{Q}) =  \sum_{\mathbf{q} \in \textrm{BZ}} \frac{\hbar S}{2}\omega_{\mathbf{q}}(\mathbf{Q}).
\label{eq:quantum}
\end{equation}  
The $ \mathcal{E}_{\text{ZP}}(\mathbf{Q})$ behaves qualitatively like the $\mathcal{A}(\mathbf{Q})$ and selects the same ordering vectors.  

For the SC lattice with $J_1=2J_2=4J_3$ the selected minima are the inequivalent points $\left(\pi,0,0\right)$, $\left(\pi,\pi,0\right)$ and $\left(\pi,\pi,\pi\right)$ and their symmetry related partners (seven points in all)~\cite{Iqbal-2016}.
 In these high symmetry points multi-$\mathbf{Q}$ states are allowed, they correspond to an ordering in real space where in a cube formed by eight neighboring lattice sites the spins sum up to zero, and this cube is repeated through the whole lattice (the magnetic superlattice is SC, with doubled lattice constant). We have performed low-T expansion for such ordering patterns, and found that the minima of the free energy correspond to any 8-sublattice collinear state, including the single-$\mathbf{Q}$ states with $\left<\pi,0,0\right>$, $\left<\pi,\pi,0\right>$ and $\left<\pi,\pi,\pi\right>$. In fact, more is true: {\it any} collinear ground state has exactly the same entropy in the harmonic approximation.  
 We believe that higher order corrections will split this degeneracy. 
 
For the FCC lattice with $J_1=2J_2$  the selected minima are the point $\left(\pi,\pi,\pi\right)$ and its three symmetry related partners. The multi-$\mathbf{Q}$ states correspond to an eight-sublattice order where every second neighbor spin pair is antiferromagnetically coupled, but otherwise the spins are oriented arbitrarily (the state consists of four interpenetrating antiferomagnetically ordered SC lattices). Here again entropy selects the  collinear, single-$\mathbf{Q}$ states forming the type-II AFM structure. \footnote{We note that MC simulations in Ref.~\onlinecite{finnish_93} found a type III, $\mathbf{Q}=(2\pi,\pi,0)$ order at low temperatures. These results call for further investigation.}

We may ask the question whether  $\mathcal{M}_{\textrm{GS}}$'s obtained by different $\Lambda^*$ can be continuously deformed  into each other by, e.g., including more shells. The $f(\mathbf{q})$ for $\Lambda^*=\Lambda$ is fully periodic in the reciprocal lattice, while the $f(\mathbf{q})$ in Eqs.~(\ref{eq:f_SC}) and (\ref{eq:fgfcc}) changes sign when translated by a unit reciprocal lattice vector. This even--odd property cannot be changed continuously, therefore the two solutions provide two different topological classes of $\mathcal{M}_{\textrm{GS}}$. The odd parity of $f(\mathbf{q})$ also pins the $f(\mathbf{Q})=0$ surface to the boundary of the Brillouin zone for the SC lattice (i.e., the $\left<\pi,q_1,q_2 \right>$ planes) and to the  $\left<\pi/2,q,\pi-q \right>$ lines in the case of the FCC lattice, while no such restriction exists for the even $f(\mathbf{q})$ function when $\bm{\delta}^*=\bm{0}$.
  
The topological distinction is further exemplified by the  Euler characteristics $\chi$ of these triply periodic surfaces~\cite{bubble_surfaces}. Let's focus on the FCC lattice. The $f(\mathbf{Q})=0$ surface from  Eq.~\eqref{eq:fgfcc} is homotopic to the so called Schwarz-P surface with $\chi=-4$. On the other hand, the extension of Eq.~\eqref{eq:epsfromgq} to FCC lattice will make a deformed sphere for $-12<c_0<0$, with $\chi=4$.  At $c_0=0$, a Lifshitz transition occurs \cite{lifshitz1960}, and for $0<c_0<4$ the $f_1(\mathbf{Q})=-c_0$ surface changes into a topologically different shape, homotopic to Schoen IWP, with $\chi=-12$. The surfaces in the case of the diamond lattice [\onlinecite{bergmann07}] belong to this latter class.

To conclude, we have provided a recipe to construct classical Heisenberg models on Bravais lattices having degenerate ground state manifolds consisting of spin-spirals. As opposed to non-Bravais lattices, the models are fine tuned, with none, or few free parameters. In their simplest case, these models can be written as the sums of interacting spins on complete graphs, providing a natural explanation for the large degeneracy of their ground state manifolds. Both thermal and quantum fluctuations  select collinear states in the semiclassical, $S \gg 1$ limit. We also show that the ground state manifolds are topologically distinct and can be classified by their Euler characteristics, however, at finite-temperatures one may expect different types of defects to appear. 
The models proposed herein can potentially serve as testbeds for future analytical and numerical studies aimed at investigating the effects of strong frustration on the critical behavior and the universality class of phase transitions, which remains to a large degree {\it terra incognita}. 
It also opens new avenues towards the realization of quantum spin liquids as has been pointed out in a recent work~\cite{niggemann_arxiv}, and it will be a worthwhile endeavor to employ state-of-the-art numerical methods to the corresponding models with small spin-$S$ to uncover possible existence of quantum spin liquid regimes.

\begin{acknowledgments}

We thank  A. S\"ut\H o for introducing us with the concept of the affine lattice.
This work was supported by the Hungarian NKFIH Grant No. K 124176 and the BME - Nanonotechnology and Materials Science FIKP grant of EMMI (BME FIKP-NAT).

\end{acknowledgments}

\bibliographystyle{apsrev4-1}
\bibliography{notes_fcc_bib} 
\end{document}